\documentclass[aps,twocolumn,prl]{revtex4-2}%
\usepackage{amsmath}
\usepackage{amssymb}
\usepackage{graphicx}
\usepackage{amsfonts}
\usepackage{array}
\usepackage{mathrsfs}
\usepackage{subfigure,overpic}
\usepackage{dcolumn}
\usepackage{epstopdf}
\usepackage{braket}
\usepackage[colorlinks,linkcolor=blue, urlcolor=blue,
anchorcolor=blue,citecolor=blue,breaklinks=true]{hyperref}
\usepackage{amsfonts}%
\providecommand{\U}[1]{\protect\rule{.1in}{.1in}}
\begin{document}

\title{Nonlinear Layer Hall Effect and Detection of the Hidden Berry Curvature Dipole in $\mathcal{PT}$-Symmetric Antiferromagnetic Insulators}
\author{Zhuo-Hua Chen}
\author{Hou-Jian Duan}
\author{Ming-Xun Deng}
\email{dengmingxun@scnu.edu.cn}
\author{Rui-Qiang Wang}
\email{wangruiqiang@m.scnu.edu.cn}
\affiliation{$^{1}$Guangdong Basic Research Center of Excellence for Structure and
Fundamental Interactions of Matter, Guangdong Provincial Key Laboratory of
Quantum Engineering and Quantum Materials, School of Physics, South China
Normal University, Guangzhou 510006, China}
\affiliation{$^{2}$Guangdong-Hong Kong Joint Laboratory of Quantum Matter, Frontier
Research Institute for Physics, South China Normal University, Guangzhou
510006, China}

\begin{abstract}
Recent experimental and theoretical studies have revealed the emergence of a
linear layer Hall effect (LHE) induced by hidden Berry curvature in \textrm{MnBi}$_{2}$\textrm{Te}$_{4}$ thin films. This phenomenon underscores the layer degree of freedom as a novel mechanism for generating Hall transport in layered materials, providing a new pathway to probe and manipulate the internal structure of fully compensated topological
antiferromagnets (AFMs). In this work, we predict a nonlinear LHE in $\mathcal{PT}$-symmetric layered AFMs, which manifests as a detectable nonlinear Hall conductivity even with respect to the AFM order and odd with respect to the vertical electric field, in contrast to the linear LHE. Furthermore, we demonstrate that the nonlinear Hall currents induced by the hidden BCD and quantum metric dipole (QMD) obey distinct symmetries and flow in different directions. Our proposed nonlinear LHE establishes an experimentally
advantageous framework for exclusively probing the hidden BCD quantum geometry.
\end{abstract}
\maketitle

Berry curvature (BC), which describes the geometry properties of Bloch bands,
plays an essential role in the development of modern topological
physics\cite{RevModPhys.82.1959}. The BC can give rise to the Hall effect when
time-reversal ($\mathcal{T}$) symmetry is globally broken. Nevertheless,
Sodemann and Fu recently proposed that nonlinear Hall effect can occur under a
second-order electric field even in $\mathcal{T}$-invariant
materials\cite{PhysRevLett.115.216806}, where breaking $\mathcal{T}$ symmetry
is no longer necessary. Instead, inversion ($\mathcal{P}$) symmetry breaking
is required. The underlying mechanism involves the dipole moment of the BC
over the occupied states, known as the Berry curvature dipole (BCD). This
innovative concept has garnered considerable
attention\cite{PhysRevB.97.041101,PhysRevLett.121.266601,Xu:2018aa,PhysRevLett.123.036806,Lee:2017aa,He:2021ab,Duan:2023aa,Sinha:2022aa,PhysRevLett.131.066301,Ho:2021aa,Kumar:2021aa,10.1093/nsr/nwac232,He:2022aa,Kang:2019aa}
and the related nonlinear transport provides a powerful tool for probing
topological physics in
solids\cite{Yasuda:2020aa,Zhao:2020aa,doi:10.1126/sciadv.aay2497,Li:2024aa,Zhang:2022aa,10.1063/5.0202692,PhysRevB.110.035139,PhysRevLett.129.227401}%
.

The BCD and the resulting nonlinear Hall effect were theoretically predicted
in bilayer \textrm{WTe}$_{\mathrm{2}}$\cite{PhysRevLett.121.266601} and later
experimentally observed\cite{Xu:2018aa}. Around the same time, the BCD has
also been detected experimentally in other materials, including
two-dimensional (2D) \textrm{MoS}$_{2}$%
\cite{Lee:2017aa,PhysRevLett.123.036806} and \textrm{WSe}$_{2}$%
\cite{10.1093/nsr/nwac232}, bilayer graphene\cite{Ho:2021aa}, Weyl
semimetals\cite{Kumar:2021aa}, and various topological
materials\cite{He:2021aa,doi:10.1073/pnas.2013386118,Tiwari:2021aa}. Even so,
probing the BCD quantum geometry is still significantly constrained by
stringent symmetry requirements\cite{PhysRevLett.115.216806}. For instance, in
$\mathcal{P}$-symmetric Weyl semimetals, the contributions from paired Weyl
nodes are the same in magnitude but opposite in sign, leading to cancellation
when summed over all Weyl nodes\cite{PhysRevB.109.155154}. Typically,
additional material engineering, such as lattice strain or interlayer
twisting\cite{PhysRevLett.123.036806,Lee:2017aa,He:2021ab,Duan:2023aa,Sinha:2022aa,PhysRevLett.131.066301}%
, is required to generate a measurable BCD. As such, novel detection and
manipulation schemes are highly desirable.

The recent discovery of 2D van der Waals antiferromagnets (AFMs), in which
both $\mathcal{P}$ and $\mathcal{T}$ symmetries are broken but $\mathcal{PT}$
symmetry is preserved\cite{PhysRevLett.124.067203}, provides a promising
platform for exploring quantum-geometry-induced nonlinear effects. The
$\mathcal{PT}$ symmetry of AFMs ensures the vanishing of global BC. However,
in A-type layered AFMs, such as \textrm{MnBi}$_{2}$\textrm{Te}$_{4}$, local BC
remains finite\cite{10.1093/nsr/nwac140}. This local BC has equal magnitude
but opposite signs in layers related by $\mathcal{PT}$ symmetry, a phenomenon
referred to as hidden BC\cite{10.1093/nsr/nwac140}. Although hidden, the local
BC can still be detected through a novel effect called the layer Hall effect
(LHE)\cite{10.1093/nsr/nwac140,PhysRevLett.122.206401,PhysRevLett.124.136407,PhysRevLett.126.156601,PhysRevB.106.245425,PhysRevB.109.115301,Gao:2021aa,Feng:2023aa,Chen:2024aa}%
, where electrons in $\mathcal{PT}$-symmetry-connected layers deflect in
opposite directions. The linear LHE has been experimentally observed in
\textrm{MnBi}$_{2}$\textrm{Te}$_{4}$\cite{Gao:2021aa}, with a remarkable value
reaching about $0.5e^{2}/h$. Based on the linear LHE, the N\'{e}el vector in
centrosymmetric magnetoelectric AFMs has been successfully
detected\cite{PhysRevLett.133.096803}, and the layer Nernst and thermal Hall
effects in 2D AFMs have also been proposed\cite{PhysRevB.111.104405}.

A natural question that arises is whether the BCD quantum geometry can
be probed in the nonlinear counterpart of the LHE. In this work, we answer
this question in the affirmative. We propose a nonlinear LHE in
$\mathcal{PT}$-symmetric AFMs induced by hidden BCD. To date, most studies of the nonlinear Hall effect in $\mathcal{PT}$-symmetric AFMs have focused on another quantum geometry, i.e., the quantum metric dipole
(QMD)\cite{Lai:2021aa,Provost:1980aa,PhysRevB.109.085419,Wang:2023aa,doi:10.1126/science.adf1506}%
. While nonlinear responses have been experimentally measured in AFMs as
probes for the QMD\cite{Wang:2023aa,doi:10.1126/science.adf1506}, the BCD
effect was overlooked due to the $\mathcal{PT}$ symmetry. Actually, although
the $\mathcal{PT}$ symmetry of AFMs results in the absence of global BCD, a
layer-locked hidden BCD exists, which has yet to be explored. It is particularly important to understand how to disentangle the BCD from QMD
quantum geometries in experiments, for the fact that when a vertical electric
field is applied, as commonly done in
experiments\cite{Wang:2023aa,doi:10.1126/science.adf1506}, the broken
$\mathcal{PT}$ symmetry allows the hidden BCD signal to coexist with the QMD.
We apply the nonlinear LHE to detect the hidden BCD
unambiguously. The BCD and QMD can be conveniently distinguished through the
nonlinear anomalous Hall conductivity, as, in certain directions, only one
type of the nonlinear Hall effect survives, and their parities in the vertical
electric field are opposite. Our findings not only expand the Hall effect
family but also provide novel proposals for quantum geometry detection.

\begin{figure}[ptb]
\centering\includegraphics[width=0.48\textwidth]{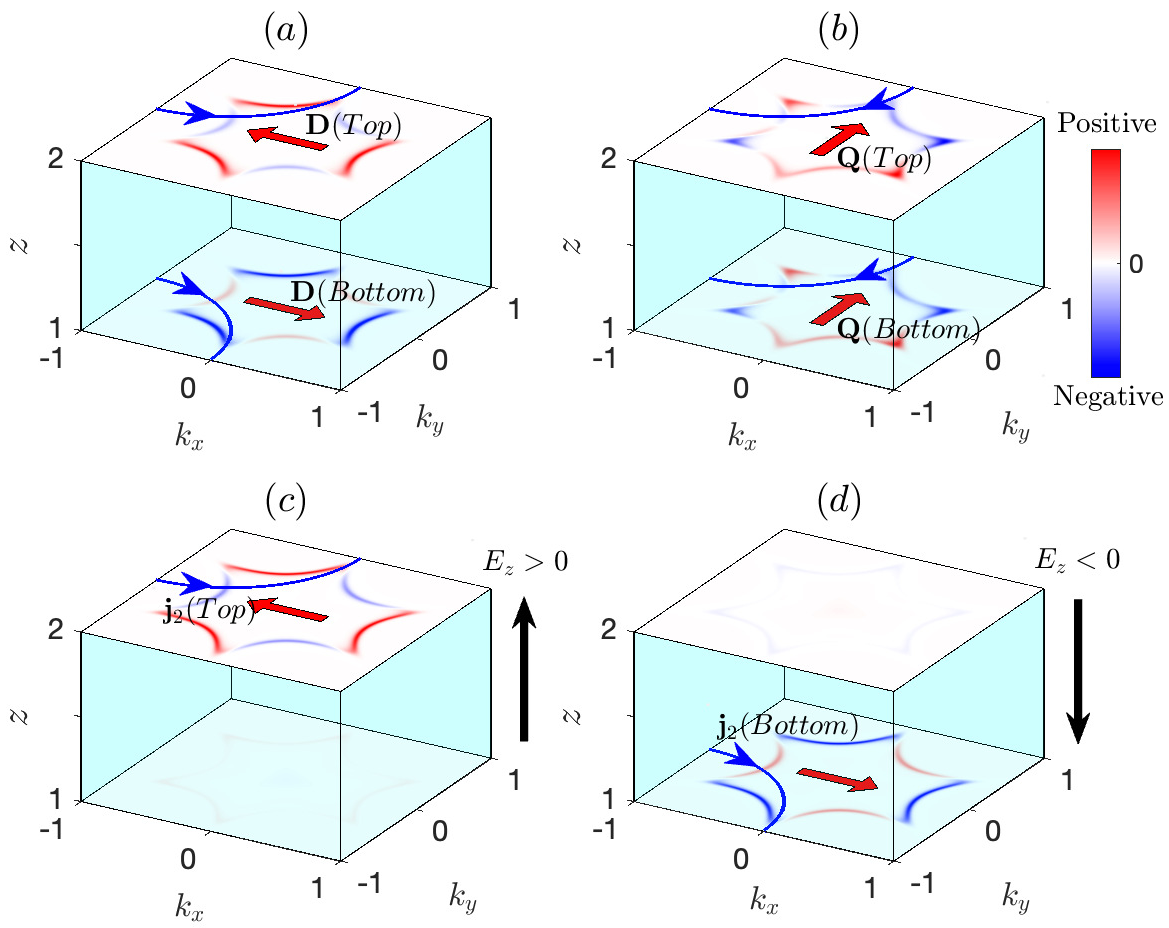}\caption{Schematics
of the (a) hidden-BCD-induced nonlinear LHE, (b) QMD-induced nonlinear
anomalous Hall effect, and (c)-(d) BCD-induced layer-polarized nonlinear Hall
currents in the presence of a vertical electric field $E_{z}$ (black arrows).
The blue and red filled hexagons indicate the distribution of the BC
$\Omega_{n}^{xy}(\boldsymbol{k},z)$ and quantum metric $\mathcal{R}_{n}%
^{yx}(\boldsymbol{k},z)$ on the Fermi surfaces, with their dipoles
$\boldsymbol{D}(z)=D_{x}^{xy}(z)\hat{x}$ and $\boldsymbol{Q}(z)=Q_{y}%
^{yx}(z)\hat{y}$ denoted by the red arrows inside the Fermi surfaces. The
resulting nonlinear Hall currents $\boldsymbol{j}_{2}(z)$ are represented by
the blue arrowed curves. In the nonlinear LHE, electrons on different layers
are deflected spontaneously to opposite directions, due to the layer-locked
BCD, while no such locking exists for the QMD.}%
\label{Fig1}%
\end{figure}

\emph{Symmetry analysis}- The concept of hidden polarization has been proposed
and applied to BC in several works
\cite{Zhang:2014aa,PhysRevLett.125.216404,PhysRevLett.118.086402,PhysRevLett.121.186401,PhysRevLett.129.276601}%
. A paradoxical duality is now recognized, where localized symmetry-breaking
mechanism governs observable physical phenomena, despite the presence of a
higher global symmetry that seemingly forbids such effects. Consider a
$\mathcal{PT}$ symmetric layered AFM stacked along the $z$-direction. The
layer-resolved BCD is defined as $\boldsymbol{D}(z)=\int d\boldsymbol{k}%
f_{\boldsymbol{k}}\partial_{\boldsymbol{k}}\Omega_{\boldsymbol{k}}(z)$, where
$\Omega_{\boldsymbol{k}}(z)$ represents the BC in the $z$-layer, and
$f_{\boldsymbol{k}}$ is the Fermi-Dirac distribution function. In A-type AFMs,
each layer is a 2D ferromagnet. The $\mathcal{PT}$ symmetry requires
$\Omega_{\boldsymbol{k}}(z)=\mathcal{PT}\Omega_{\boldsymbol{k}}(z)\mathcal{T}%
^{-1}\mathcal{P}^{-1}=-\Omega_{\boldsymbol{k}}(\bar{z})$, where $\bar
{z}=\mathcal{P}z\mathcal{P}^{-1}$ is the inversion partner of $z$. The
$\mathcal{PT}$-odd BC is fundamental to the linear LHE. Because $\mathcal{PT}%
\partial_{\boldsymbol{k}}\mathcal{T}^{-1}\mathcal{P}^{-1}=$\ $\partial
_{\boldsymbol{k}}$ is $\mathcal{PT}$-even, the momentum-resolved BCD
$\boldsymbol{D}(\boldsymbol{k},z)=\partial_{\boldsymbol{k}}\Omega
_{\boldsymbol{k}}(z)$ is $\mathcal{PT}$-odd, resulting in $\boldsymbol{D}%
(\boldsymbol{k},z)=-\boldsymbol{D}(\boldsymbol{k},\bar{z})$. Although the
global $\mathcal{PT}$ symmetry ensures that the net BCD vanishes,
$\boldsymbol{D}(z)$ can exhibit a nonzero distribution across different
layers, which is the so-called hidden BCD.

Naturally, upon the application of an electric field $\boldsymbol{E}$, the
nonlinear Hall currents induced by the hidden BCD, $\boldsymbol{j}%
_{2}(z)=\frac{e^{3}\tau}{2}\hat{z}\times\boldsymbol{E}[\boldsymbol{D}%
(z)\cdot\boldsymbol{E}]$, as derived in Ref. \cite{PhysRevLett.115.216806},
are equal in magnitude but opposite in sign for the $z$- and $\bar{z}$-layers.
This implies that the electrons on the $\mathcal{PT}$-symmetry-related layers
will deflect spontaneously in opposite directions, as depicted in Fig.
\ref{Fig1}(a). Since the LHE is contributed by nonlinear Hall currents, we dub
it the nonlinear LHE. Note that nonlinear Hall effects can also arise from the
nonlinear Drude and QMD mechanisms. However, both the Drude and QMD components
are $\mathcal{PT}$-even and thus do not contribute to the nonlinear LHE, as
illustrated in Fig. \ref{Fig1}(b). Therefore, only the BCD mechanism needs to
be considered when discussing the nonlinear LHE.

\emph{Theoretical model}- Since \textrm{MnBi}$_{2}$\textrm{Te}$_{4}$ is
extensively employed experimentally, we start from its low-energy effective
Hamiltonian, whose nonmagnetic part reads
\cite{10.1093/nsr/nwac140,PhysRevLett.122.206401,PhysRevLett.124.136407,PhysRevB.106.245425,PhysRevLett.126.156601,PhysRevB.109.115301}%
\begin{equation}
\mathcal{H}\left(  \boldsymbol{k}\right)  =\left(  \boldsymbol{d}%
_{\boldsymbol{k}}\cdot\boldsymbol{s}\right)  \sigma_{x}+M_{\boldsymbol{k}%
}\sigma_{z}+w\left(  k_{+}^{3}+k_{-}^{3}\right)  \sigma_{y}, \label{eq_1}%
\end{equation}
where $\boldsymbol{d}_{\boldsymbol{k}}=\left(  A_{1}k_{x},A_{1}k_{y}%
,A_{2}k_{z}\right)  $, $k_{\pm}=k_{x}\pm ik_{y}$, $M_{\boldsymbol{k}}%
=M_{0}-B_{1}k_{z}^{2}-B_{2}(k_{x}^{2}+k_{y}^{2})$, and $\boldsymbol{s}$
($\boldsymbol{\sigma}$) represents the vector of the spin (orbit) Pauli
matrices. Here, the $\mathcal{P}$ and $\mathcal{T}$ operators are given by
$\mathcal{P}=\sigma_{z}$ and $\mathcal{T}=is_{y}\mathcal{K}$, respectively,
with $\mathcal{K}$ being the complex conjugation operator. The model
parameters can be found in Ref. \cite{10.1093/nsr/nwac140}. The case with
$M_{0}/B_{1}>0$ describes a 3D topological insulator, and the hexagonal
warping term $\sim w$ reduces the full rotation symmetry down to $C_{3z}$. The
A-type AFM order and the vertical electric field along the stacking $z$
direction can be included through the tight-binding Hamiltonian%
\begin{equation}
H=\sum_{z=1}^{n_{z}}c_{z}^{\dag}\left[  \left(  h_{0}+V_{z}\right)
c_{z}+h_{1}c_{z+1}+h_{-1}c_{z-1}\right]  , \label{eq_Hz}%
\end{equation}
with $V_{z}=(-1)^{z}m\boldsymbol{N}\cdot\boldsymbol{s}+eE_{z}[z-(n_{z}+1)/2]$
and $h_{n}=\int dk_{z}\mathcal{\tilde{H}}\left(  \boldsymbol{k}\right)
e^{ink_{z}}$. Here, $\boldsymbol{N}=(\sin\theta\cos\phi,\sin\theta\sin
\phi,\cos\theta)$ represents the N\'{e}el vector of the AFM, and
$\mathcal{\tilde{H}}\left(  \boldsymbol{k}\right)  $ corresponds to Eq.
(\ref{eq_1}) with the transformations $k_{z}\rightarrow\sin k_{z}$ and
$k_{z}^{2}\rightarrow2-2\cos k_{z}$, as described in Refs.
\cite{PhysRevB.109.125111,PhysRevB.110.075136}.\begin{figure}[ptb]
\centering\includegraphics[width=0.48\textwidth]{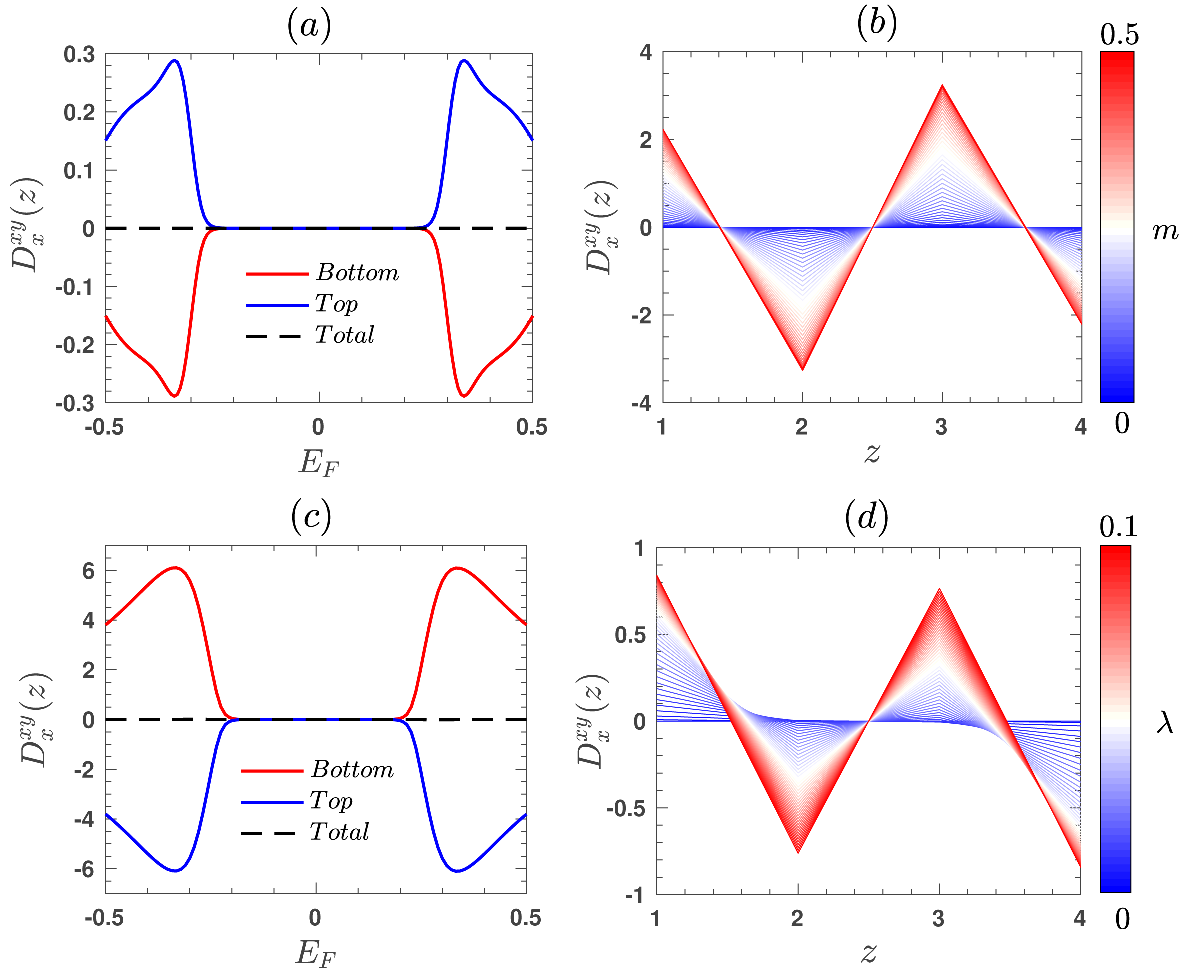}
\caption{Numerical results for the layer-resolved BCD induced by (a)-(b) the
AFM order with $m=0.5$, and (c)-(d) the tilt with $\lambda=0.1$, where
$n_{z}=2$ and $n_{z}=4$ are chosen in the left and right panels, respectively.
The rest parameters for the calculations are taken as $\theta=0$, $\phi=0$,
$w=0.5$, $M_{0}=0.1$, $A_{1}=A_{2}=0.55$, $B_{1}=B_{2}=0.25$, and $E_{z}=0$.}%
\label{Fig2}%
\end{figure}

The layer-resolved nonlinear conductivity induced by the BCD can be expressed
as\cite{PhysRevLett.132.026301,PhysRevB.108.L241104}%
\begin{equation}
\sigma_{\gamma;\alpha\beta}^{BCD}(z)=-\frac{e^{3}\tau}{\hbar^{2}}[D_{\alpha
}^{\beta\gamma}(z)+D_{\beta}^{\alpha\gamma}(z)], \label{eq_BCD}%
\end{equation}
where $D_{\alpha}^{\beta\gamma}(z)=\sum_{n}\int d\boldsymbol{k}f_{n}%
\partial_{k_{\alpha}}\Omega_{n}^{\beta\gamma}(\boldsymbol{k},z)$ is the
$\alpha$-th component of the BCD vector, and $f_{n}=1/\left[  e^{\left(
\varepsilon_{n}-E_{F}\right)  /k_{B}T}+1\right]  $ denotes the equilibrium
distribution function. The layer-resolved BC corresponds to the imaginary part
of the quantum geometric tensor\cite{PhysRevB.88.064304}
\begin{equation}
\mathcal{Q}_{mn}^{\beta\gamma}(\boldsymbol{k},z)=\frac{\langle m|\upsilon
_{\beta}|n\rangle\langle n|\upsilon_{\gamma}(z)|m\rangle}{\left(
\varepsilon_{m}-\varepsilon_{n}\right)  ^{2}}, \label{eq_BC}%
\end{equation}
namely, $\Omega_{n}^{\beta\gamma}(\boldsymbol{k},z)=\sum_{m\neq n}%
2\operatorname{Im}[\mathcal{Q}_{mn}^{\beta\gamma}(\boldsymbol{k},z)]$. Here,
$\left\vert n\right\rangle $ is wavefunction for the energy $\varepsilon_{n}$
of Eq. (\ref{eq_Hz}), $\upsilon_{\beta}=\partial_{k_{\beta}}\mathcal{H}\left(
\boldsymbol{k}\right)  $ and $\upsilon_{\gamma}(z)=\hat{P}_{z}\upsilon
_{\gamma}\hat{P}_{z}$, in which $\hat{P}_{z}$ denotes the projection operator
constructed by the wavefunction on the $z$-layer
\cite{PhysRevB.92.081107,Xu:2021aa,PhysRevResearch.5.013001}. From Eq.
(\ref{eq_BC}), we can define the layer-resolved band-normalized quantum metric
$\mathcal{R}_{n}^{\beta\gamma}(\boldsymbol{k},z)=\sum_{m\neq n}%
\operatorname{Re}[\mathcal{Q}_{mn}^{\beta\gamma}(\boldsymbol{k},z)/\left(
\varepsilon_{m}-\varepsilon_{n}\right)  ]$. The layer-resolved QMD-induced
nonlinear conductivity takes the
form\cite{PhysRevLett.132.026301,PhysRevB.108.L241104}
\begin{equation}
\sigma_{\gamma;\alpha\beta}^{QMD}(z)=-\frac{e^{3}}{\hbar}\{2Q_{\gamma}%
^{\alpha\beta}(z)-\frac{1}{2}[Q_{\alpha}^{\beta\gamma}(z)+Q_{\beta}%
^{\alpha\gamma}(z)]\}, \label{eq_QMD}%
\end{equation}
with $Q_{\alpha}^{\beta\gamma}(z)=\sum_{n}\int d\boldsymbol{k}f_{n}%
\partial_{k_{\alpha}}\mathcal{R}_{n}^{\beta\gamma}(\boldsymbol{k},z)$. The
$\mathcal{PT}$ symmetry ensures
\begin{equation}
\mathcal{Q}_{mn}^{\beta\gamma}(\boldsymbol{k},z)=\mathcal{PTQ}_{mn}%
^{\beta\gamma}(\boldsymbol{k},z)\mathcal{T}^{-1}\mathcal{P}^{-1}%
=[\mathcal{Q}_{mn}^{\beta\gamma}(\boldsymbol{k},\bar{z})]^{\ast},
\end{equation}
such that%
\begin{equation}
\mathcal{R}_{n}^{\beta\gamma}(\boldsymbol{k},z)=\mathcal{R}_{n}^{\beta\gamma
}(\boldsymbol{k},\bar{z})\text{, }\Omega_{n}^{\beta\gamma}(\boldsymbol{k}%
,z)=-\Omega_{n}^{\beta\gamma}(\boldsymbol{k},\bar{z}).
\end{equation}
Consequently, the $\mathcal{PT}$-odd hidden BCD contributes to the nonlinear
LHE, while the $\mathcal{PT}$-even QMD makes no contribution.

In the AFM \textrm{MnBi}$_{2}$\textrm{Te}$_{4}$, the $\mathcal{PT}$-symmetric
staggered magnetization breaks the interlayer $\mathcal{P}$ symmetry and
intralayer $\mathcal{T}$ symmetry, leading to $\Omega_{n}^{\beta\gamma
}(\boldsymbol{k},z)=-\Omega_{n}^{\beta\gamma}(-\boldsymbol{k},\bar{z})$ and
$\Omega_{n}^{\beta\gamma}(\boldsymbol{k},z)=\Omega_{n}^{\beta\gamma
}(-\boldsymbol{k},z)$ when $w=0$. As a result, the BC with only out-of-plane
component behaves as a pseudoscalar globally and the BCD in each layer behaves
as a pseudovector. For $m=0$, the warping term with both $\mathcal{P}$ and
$\mathcal{T}$ symmetries can also induce a BC that fulfills $\Omega_{n}%
^{\beta\gamma}(\boldsymbol{k},z)=\Omega_{n}^{\beta\gamma}(-\boldsymbol{k}%
,\bar{z})$ and $\Omega_{n}^{\beta\gamma}(\boldsymbol{k},z)=-\Omega_{n}%
^{\beta\gamma}(-\boldsymbol{k},z)$. Consequently, when both $m$ and $w$ are
finite, the parity of the BC with respect to $\boldsymbol{k}$ is broken.
Nevertheless, the $C_{3z}$ symmetry with a mirror line along the $x$-axis
ensures $\mathcal{M}_{y}\Omega_{n}^{\beta\gamma}(\boldsymbol{k},z)\mathcal{M}%
_{y}^{-1}=\Omega_{n}^{\beta\gamma}(\boldsymbol{k},z)$ with $\mathcal{M}%
_{y}k_{y}\mathcal{M}_{y}^{-1}=-k_{y}$, as seen from Fig. \ref{Fig1}(a).
Therefore, $\mathcal{M}_{y}\partial_{k_{y}}\Omega_{n}^{\beta\gamma
}(\boldsymbol{k},z)\mathcal{M}_{y}^{-1}=-\partial_{k_{y}}\Omega_{n}%
^{\beta\gamma}(\boldsymbol{k},z)$, making $D_{y}^{\beta\gamma}(z)$ vanish. The
combined effects of the staggered magnetization and warping term give rise to
a nonzero layer-locked BCD $D_{x}^{\beta\gamma}(z)$. Since $\boldsymbol{j}%
_{2}\left(  z\right)  \propto\hat{z}\times\boldsymbol{E}[\boldsymbol{D}%
(z)\cdot\boldsymbol{E}]$, the nonlinear LHE arises when the electric
field is exerted along the $x$ direction, with the layer Hall current
orthogonal to the BCD vector $\boldsymbol{D}(z)=D_{x}^{xy}(z)\hat{x}$. 

\begin{figure}[ptb]
\centering
\includegraphics[width=0.48\textwidth]{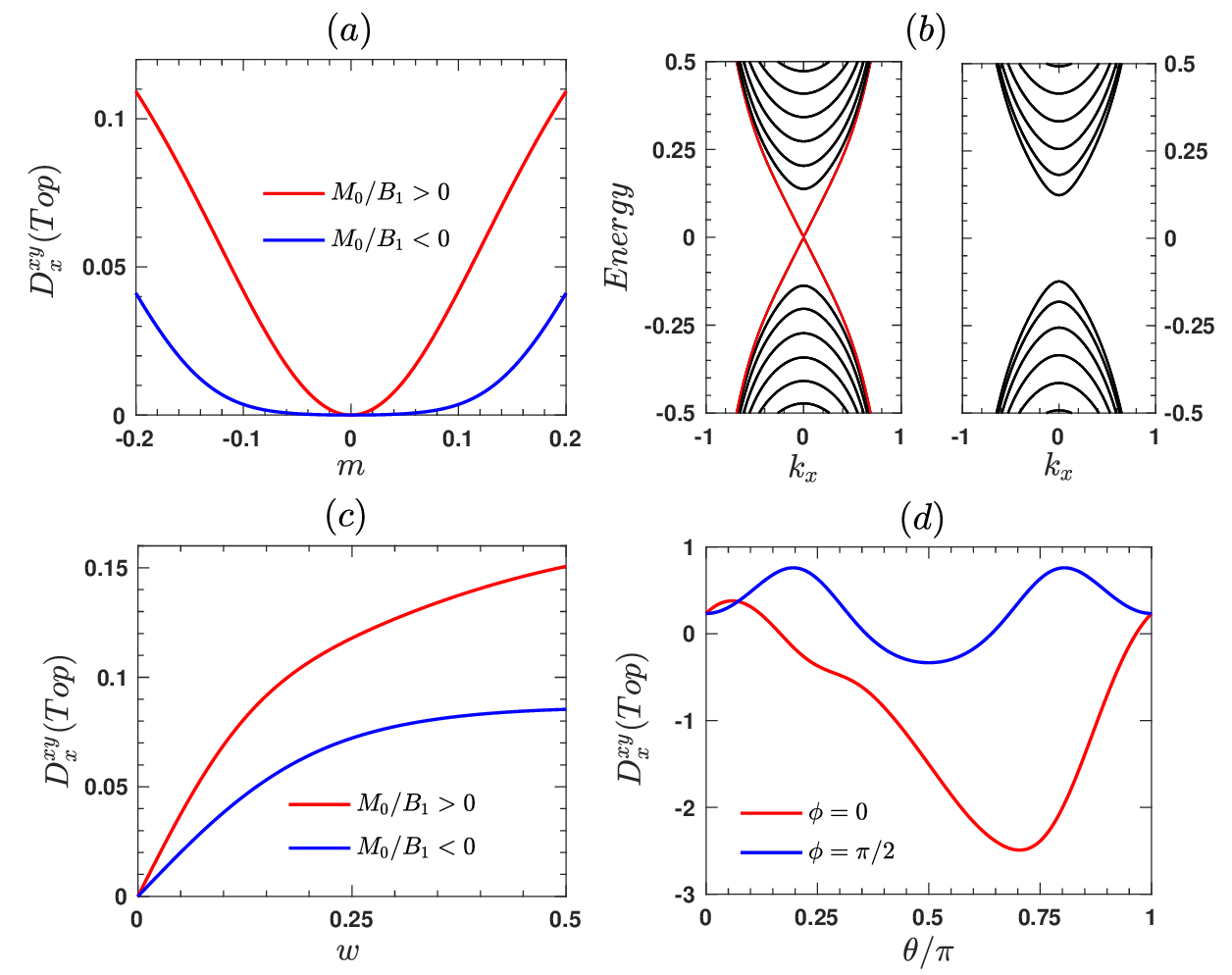} \caption{(a) The
layer-resolved nonlinear Hall conductivity $\sigma_{y;xx}^{BCD}(Top)=-\sigma
_{y;xx}^{BCD}(Bottom)$ as a function of $m$, with $\boldsymbol{N}=(0,0,1)$,
$|M_{0}/B_{1}|=0.4$, and $n_{z}=2$. (b) The energy band for $m=0$ and
$n_{z}=20$, with $M_{0}/B_{1}<0$ (right) and $M_{0}/B_{1}>0$ (left), where
topological surface states emerge in the bulk band gap in the phase of
topological AFM insulator. (c)-(d) $\sigma_{y;xx}^{BCD}\left(  Top\right)  $
as a function of $w$ and the direction of the N\'{e}el vector, respectively.
The rest parameters are the same as in Fig. \ref{Fig2}(a).}%
\label{Fig3}%
\end{figure}

\emph{Numerical results}- Firstly, we discuss the simplest case with the
N\'{e}el vector along the $z$-direction. The layer-resolved BCD as a function
of the Fermi energy $E_{F}$ is plotted in Fig. \ref{Fig2}(a). Obviously, for a
bilayer structure, $D_{x}^{xy}(Top)=-D_{x}^{xy}(Bottom)$ is guaranteed by the
global $\mathcal{PT}$ symmetry. For structures with more layers, the
$\mathcal{PT}$ symmetric AFM pattern is always preserved, as illustrated in
Fig. \ref{Fig2}(b), where $D_{x}^{xy}(z)=-D_{x}^{xy}(\bar{z})$ with $\bar
{z}=n_{z}+1-z$. This implies that the nonlinear Hall conductivities from
$\mathcal{PT}$ partners are exactly compensated, thereby establishing the
nonlinear LHE induced by the hidden BCD. In contrast to the linear LHE, which
originates from $m$-odd BC, the nonlinear LHE is governed by $m$-even BCD, as
shown in Fig. \ref{Fig3}(a). Accordingly, in the nonlinear LHE, it is the
breaking of $\mathcal{P}$-symmetry, rather than $\mathcal{T}$-symmetry, that
plays a crucial role. To substantiate this, we replace the magnetization
$m\boldsymbol{N}\cdot\boldsymbol{s}$ with a tilt $\lambda k_{x}$ and plot the
BCD in Figs. \ref{Fig2}(c) and (d). As expected, the scenario is reproduced
similar to Figs. \ref{Fig2}(a) and (b). This suggests that the proposed
nonlinear LHE can also be realized in more nonmagnetic layered materials.

In Fig. \ref{Fig3}, we plot the nonlinear layer Hall conductivity
$\sigma_{y;xx}^{BCD}(Top)=-\sigma_{y;xx}^{BCD}(Bottom)$ as functions of the
AFM order and warping parameters. In the phase of topological AFM insulator
with $M_{0}/B_{1}>0$, $\sigma_{y;xx}^{BCD}(z)$ can be more than one order of
magnitude greater than that in the trivial insulator with $M_{0}/B_{1}<0$, as
observed in Fig. \ref{Fig3}(a). The significant enhancement of $\sigma
_{y;xx}^{BCD}(z)$ is attributed to the topological surface states, as shown in
Fig. \ref{Fig3}(b), where the surface states within the bulk band gap become
decoupled and observable for $n_{z}>20$. Different from the linear
LHE\cite{10.1093/nsr/nwac140,PhysRevLett.122.206401,PhysRevLett.124.136407,PhysRevLett.126.156601,PhysRevB.106.245425,PhysRevB.109.115301,Gao:2021aa,Feng:2023aa,Chen:2024aa}%
, the nonlinear LHE requires the warping term to lower the full rotation
symmetry in each layer. Specifically, the interplay of the warping term and
AFM order breaks the symmetry of $D_{x}^{xy}(\boldsymbol{k},z)$ in $k_{x}$,
ensuring the nonzero result of $\sigma_{y;xx}^{BCD}(z)$. As the warping
parameter $w$ increases, the nonlinear LHE becomes more pronounced, as
indicated in Fig. \ref{Fig3}(c). Moreover, the nonlinear LHE can be further
enhanced by tuning the N\'{e}el vector from out-of-plane to in-plane, as shown
in Fig. \ref{Fig3}(d).

\begin{figure}[ptb]
\centering
\includegraphics[width=0.48\textwidth]{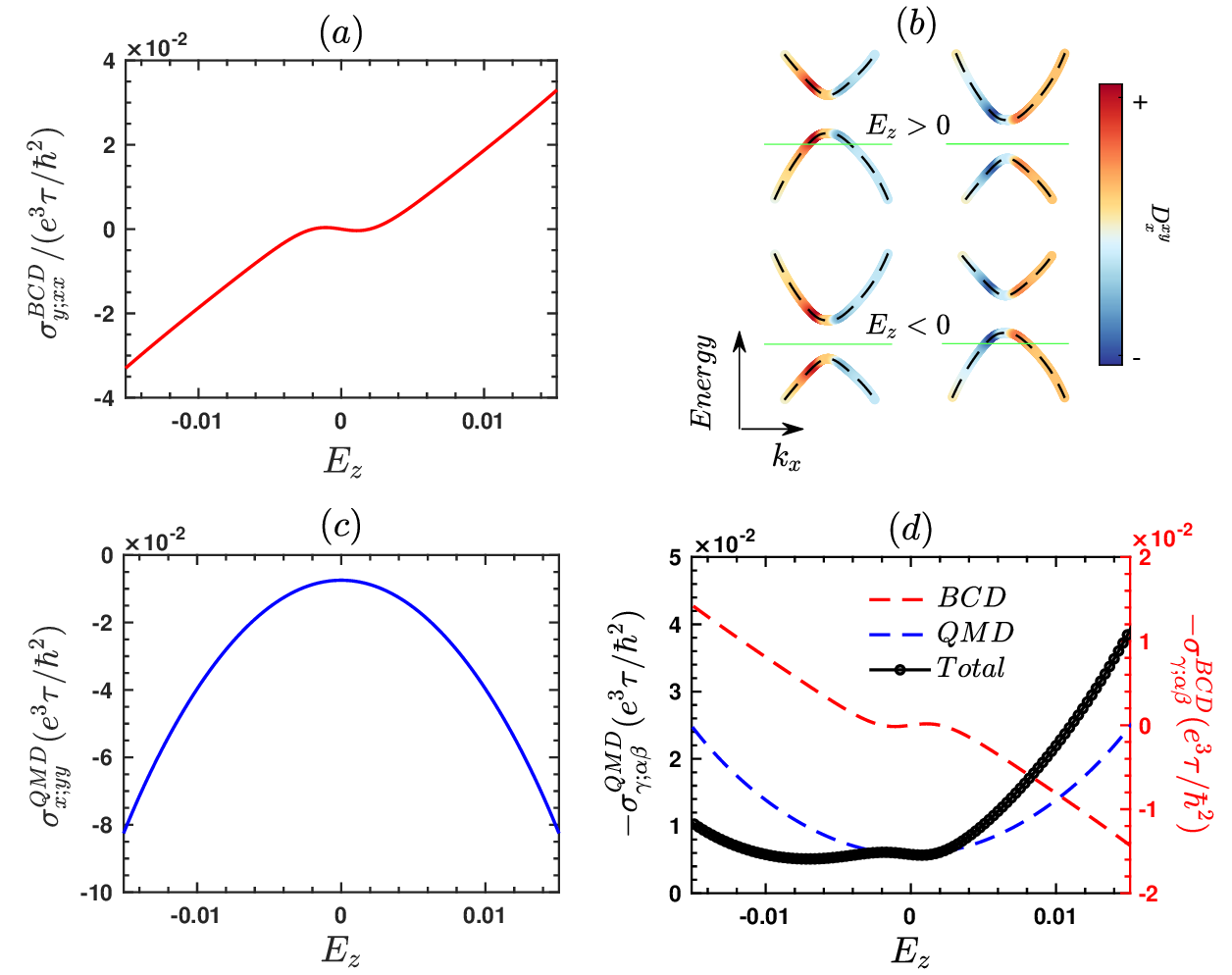} \caption{(a) The net
nonlinear Hall conductivity $\sigma_{y;xx}^{BCD}$ as a function of the
vertical electric field. (b) The underlying physics of the hidden-BCD-induced
nonlinear Hall conductivity, where the vertical electric field induces a
potential difference between the top and bottom surfaces, leading to an
imbalance of their Hall conductivity, as a result of the uncompensated hidden
BCD of different layers when the global $\mathcal{PT}$ symmetry is broken. (c)
The $E_{z}$-dependence of the QMD component $\sigma_{x;yy}^{QMD}$ and (d) the
nonlinear Hall conductivities when the driving electric field is misaligned
from the crystal axis.}%
\label{Fig4}%
\end{figure}

As discussed above, due to the $\mathcal{PT}$ symmetry, the BCDs on the $z$-
and $\bar{z}$-layers are compensated, resulting in a vanishing net nonlinear
Hall conductivity. In order to observe the nonlinear LHE, the $\mathcal{PT}$
symmetry must be broken, e.g., by applying a vertical electric field as
illustrated in Figs. \ref{Fig1}(c) and (d). Indeed, as shown in Fig.
\ref{Fig4}(a), a net anomalous nonlinear Hall conductivity emerges for finite
$E_{z}$. Importantly, the nonlinear Hall signal flips sign when reversing
$E_{z}$. This electric-field-reversible nonlinear Hall effect represents a
fundamental piece of evidence for the nonlinear LHE. Physically, this behavior
can be understood as follows. As $E_{z}$ increases, the $\mathcal{PT}$
symmetry is progressively disrupted, causing the degenerate bands to split,
see Fig. \ref{Fig4}(b). Therefore, the BCD, in bilayer \textrm{MnBi}$_{2}%
$\textrm{Te}$_{4}$ for example, is dominated by the top/bottom layer, when the
Fermi level crosses the band from the top/bottom layer. As a result, the BCDs
on the $z$- and $\bar{z}$-layers are no longer compensated, leading to a
detectable nonlinear anomalous Hall conductivity. 

Notice that recent experiments have also investigated the application of an
out-of-plane electric field in 2D materials, which can induce an electrically
switchable
BCD\cite{Xu:2018aa,PhysRevLett.102.256405,PhysRevLett.130.016301,PhysRevB.110.165401}%
. In those works, the electric field is used to tune the band gap, driving a
topological phase transition from a quantum spin Hall state to a normal
insulating state, thereby enhancing the BCD near the topological critical
points. The underlying physics of the nonlinear LHE proposed here is
fundamentally different. In our case, the vertical electric field is adopted
to lift the degeneracy imposed by $\mathcal{PT}$ symmetry, which provides an
alternative pathway to detect or manipulate the BCD and its associated
nonlinear response.

\emph{Comparison of BCD and QMD contributions}-In $\mathcal{PT}$ symmetric
AFMs, the nonlinear Hall effect can also be induced by the
QMD\cite{Lai:2021aa,Provost:1980aa,PhysRevB.109.085419,Wang:2023aa,doi:10.1126/science.adf1506}%
. However, the contribution from the QMD has the same sign across different
layers, thus not contributing to the nonlinear LHE. To demonstrate this, we
estimate the QMD contribution through Eq. (\ref{eq_QMD}), and verify that for
the Hamiltonian of \textrm{MnBi}$_{2}$\textrm{Te}$_{4}$, as given in Eq.
(\ref{eq_Hz}), $\sigma_{\gamma;\alpha\beta}^{QMD}$ vanishes whenever there is
an odd number of $y$-subscripts, since the integral kernel in $Q_{\alpha
}^{\beta\gamma}(z)$ is an odd function of $k_{y}$. As a consequence, only
$\sigma_{x;yy}^{QMD}$ survives in the nonlinear anomalous Hall conductivity.
In contrast, the dominated BCD contribution is $\sigma_{y;xx}^{BCD}$.
Physically, the $\mathcal{M}_{y}$-odd symmetry of $\mathcal{R}_{n}%
^{\beta\gamma}(\boldsymbol{k},z)$, i.e., $\mathcal{M}_{y}\mathcal{R}%
_{n}^{\beta\gamma}(\boldsymbol{k},z)\mathcal{M}_{y}^{-1}=-\mathcal{R}%
_{n}^{\beta\gamma}(\boldsymbol{k},z)$, different from the $\mathcal{M}_{y}%
$-even BC, causes the two nonlinear Hall currents to be perpendicular, as
shown in Figs. \ref{Fig1}(a) and (b). Thus, when the driving field is along a
particular crystal axis, e.g., $x$- or $y$- axis, only one type of nonlinear
Hall current survives, making it straightforward to distinguish between the
BCD and QMD contributions.

Except for the direction-locked property, the BCD and QMD contributions can
also be distinguished by their different dependences on the vertical electric
field. As shown in Fig. \ref{Fig4}(a), $\sigma_{y;xx}^{BCD}$ exhibits
antisymmetry with respect to $E_{z}$ due to the nonlinear LHE, while
$\sigma_{x;yy}^{QMD}$ is symmetric. Furthermore, $\sigma_{x;yy}^{QMD}$ can be
finite even in the absence of $E_{z}$, as illustrated in Fig. \ref{Fig4}(c),
indicating that the QMD mechanism for the nonlinear Hall effect does not
require the breaking of $\mathcal{PT}$ symmetry. When the driving electric
field is misaligned from the crystal axis, the coexistence of both mechanisms
leads to an asymmetric lineshape, as shown in Fig. \ref{Fig4}(d). In this
situation, one can isolate the symmetric component as the QMD contribution and
the antisymmetric component as the BCD contribution. Notably, recent
experiments on even-layer $\mathrm{MnBi}_{2}\mathrm{Te}_{4}$ have shown that
the measured nonlinear Hall conductivity exhibits an asymmetric dependence on
$E_{z}$, see Fig. 4(b) of Ref. \cite{Wang:2023aa} and Fig. 3(d) of Ref.
\cite{doi:10.1126/science.adf1506}.

\emph{Conclusion}-In summary, we have proposed a nonlinear LHE induced by the
hidden BCD in $\mathcal{PT}$-symmetric topological AFM insulators. The
resulting nonlinear Hall conductivity exhibits a unique direction-locked
property and a distinctive dependence on the modulation of the vertical
electric field, contrasting with the QMD contribution. This nonlinear LHE
provides an experimentally advantageous framework for probing the hidden BCD quantum geometry, even when it coexists with the QMD contribution. Moreover, the
hidden-BCD-induced nonlinear LHE, which is sensitive to the orientation of the
N\'{e}el vector, offers a means for the N\'{e}el vector detection in
magnetoelectric AFMs with both in-plane and out-of-plane N\'{e}el vectors. Our
results not only expand the Hall effect family but also extend the BCD physics
to a broad range of material platforms with high tunability, offering
promising prospects for practical applications.

This work was supported by the National NSF of China under Grants No.
12274146, No. 12174121 and No. 12104167; the Guang dong Basic and Applied
Basic Research Foundation under Grant No. 2023B1515020050; the Guang dong NSF
of China under Grant No. 2024A1515011300; and the Guangdong Provincial Quantum
Science Strategic Initiative under Grant No. GDZX2401002.

\bibliography{bibCZH2025}

\end{document}